\documentclass[rnote]{aa} 
\usepackage{graphicx}
\usepackage{txfonts}
\usepackage{natbib}
\usepackage{url}
\usepackage{multirow}
\usepackage{psfig}
\begin{document} 

   \title{Constraining Sub-Grid Physics with High-Redshift Spatially-Resolved Metallicity Distributions}

   \author{B.~K. Gibson \inst{1}
          \and K. Pilkington \inst{1}
          \and C.~B. Brook \inst{2}
          \and G.~S. Stinson \inst{3}
          \and J. Bailin \inst{4}}

   \institute{Jeremiah Horrocks Insitute, University of Central Lancashire,
              Preston, PR1~2HE, UK\\
          \email{bkgibson@uclan.ac.uk}
          \and Departamento de F\'isica Te\'orica, Universidad Aut\'onoma de
            Madrid, Cantoblanco, Madrid, E28049, Spain
          \and Max-Planck-Institut f\"ur Astronomie, K\"onigstuhl 17,
            Heidelberg, 69117, Germany
          \and Department of Physics \& Astronomy, University of Alabama, 
            Tuscaloosa, AL, 35487-0324, USA}


 
  \abstract
   {} 
   {We examine the role of energy feedback in shaping
     the distribution of metals within cosmological
     hydrodynamical simulations of L$^\ast$ disc galaxies. While
     negative abundance gradients today provide a boundary condition for
     galaxy evolution models, in support of inside-out disc growth, empirical
     evidence as to whether abundance gradients steepen or flatten with
     time remains highly contradictory.}
   {We made use of a suite of L$^\ast$ discs, realised with and without
     `enhanced' feedback.  All the simulations were produced using
     the smoothed particle hydrodynamics code {\tt Gasoline\rm}, and their
     in situ gas-phase metallicity gradients traced from
     redshift $z$$\sim$2 to the present-day.  Present-day age-metallicity
     relations and metallicity distribution functions were derived
     for each system.}
   {The `enhanced' feedback models, which have been shown to be in 
     agreement with a broad range of empirical scaling relations, 
     distribute energy and re-cycled ISM material over large scales
     and predict the existence of relatively `flat' and temporally invariant 
     abundance gradients.  Enhanced feedback schemes reduce significantly
     the scatter in the local stellar age-metallicity relation and, 
     especially, the [O/Fe]-[Fe/H] relation.  The local [O/Fe] distribution 
     functions for our L$^\ast$ discs show clear bimodality, with peaks 
     at [O/Fe]=$-$0.05 and $+$0.05 (for stars with [Fe/H]$>$$-$1), 
     consistent with our earlier work on dwarf discs.}
   {Our results with `enhanced' feedback are
     inconsistent with our earlier generation of 
     simulations realised with `conservative' feedback. We conclude that 
     spatially-resolved metallicity distributions, particularly at 
     high-redshift, offer a unique and under-utilised constraint on
     the uncertain nature of stellar feedback processes.}
    
   \keywords{galaxies: abundances -- galaxies: evolution -- galaxies: formation -- Galaxy: disc}

   \maketitle
%

\section{Introduction}
Radial abundance gradients and stellar age-metallicity relations provide 
two powerful constraints on the complex (and poorly understood) 
interplay between gas infall (e.g., cold flows from the intergalactic 
medium, coronal re-cycling of the underlying ISM), outflows (e.g., 
galactic fountains, superwinds, mass loading), stellar migration and 
radial gas flows, secular kinematic heating, interaction- and 
merger-driven energetics, and star formation efficiency, in driving the 
`inside-out' growth of disc galaxies.  Local discs, including the Milky 
Way, provide one critical `boundary condition' for all models, in the 
sense that their present-day (i.e., gas-phase) radial metallicity 
gradients must be `negative' (i.e., decreasing in metallicity with 
increasing galactocentric radius) and of the order $-$0.04~dex/kpc.

Prior to 2011, no in situ measurements of abundance gradients 
at $z$>0 existed (particularly for `typical' star-forming and/or Grand 
Design spirals); three datasets have started to change this picture. 
MASSIV \citep{Quey12} found essentially flat gradients in a large number of 
discs (both isolated and interacting) at $z$$\sim$1, although
the challenging nature of this non-adaptive optics work makes the
results particularly sensitive to spatial resolution limitations
\citep{Yuan13}. Conversely, 
\citet{Yuan11} and \citet{Jone12}, using reconstructed source plane 
images of gravitationally-lensed discs at 1.5$<$$z$$<$2.5, found that in 
three out of their four systems, the inferred oxygen gradients were very 
steep ($-$0.15$\rightarrow$$-$0.3~dex/kpc).

Complicating the picture further, one {\it could \rm} try and make use 
of planetary nebulae sub-types within the Milky Way; as planetaries 
arise from stars ranging from $\sim$1~M$_\odot$ to $\sim$8~M$_\odot$, 
they provide something of a temporal probe of abundance gradients, 
albeit not measured in situ, but instead measured today at 
$z$=0, after experiencing several Gyrs or more of potential kinematic 
heating (both random `blurring' and systematic radial `migration'). 
Extant attempts to infer the temporal evolution of the Milky Way's 
gradient using such planetaries \citep{Maci03,Stan10} reflects that 
there remains a significant discrepancy between the claimed 
behaviour;\footnote{Very uncertain distance determinations and nebular 
emission approaches, presumably cloud the issue, but this is beyond the 
scope of our abilities to disentangle.} The \citet{Maci03} analysis 
results in an inferred gradient for the Milky Way which was steep at 
early times and flattened to today's value ($-$0.04~dex/kpc: 
\citealt{Rupk10}). Conversely, the \citet{Stan10} work leads to an 
inferred gradient at early times which is somewhat flatter than today's 
value (hence, a steepening with time).

Motivated by these empirical constraints, we examine here the 
role of energy feedback in shaping the temporal evolution of abundance 
gradients and age-metallicity relations within a sub-set of cosmological 
hydrodynamical disc simulations drawn from the MUGS
(McMaster Unbiased Galaxy Simulations: \citealt{Stin10}) and 
MaGICC (Making Galaxies in a Cosmological Context: \citealt{Broo12b}) 
suites. We will demonstrate how such observations 
can genuinely constrain the highly uncertain nature and magnitude of 
energy feedback underpinning galaxy formation.

In \citet{Pilk12a}, we showed that `conventional' feedback schemes 
(i.e., those making use of $\sim$10-40\% of the energy associated with 
each supernova (SN), to heat the surrounding ISM), when 
coupled with a classical SPH 
approach to hydrodynamics (independent of the SPH code employed), tended 
to result in galaxies with steep abundance gradients at redshifts 
$z$$>$1 ($-$0.15$\rightarrow$$-$0.30~dex/kpc vs $-$0.04~dex/kpc, today). 
Such `conventional' feedback schemes, when coupled to a grid-based 
approach to hydrodynamics {\it at roughly the same resolution\rm}, also 
led to gradients steeper at high-redshift relative to the predicted 
present-day values ($-$0.05$\rightarrow$$-$0.10~dex/kpc vs 
$-$0.04~dex/kpc, today), although systematically shallower than their 
SPH counterparts.  This systematic difference was driven by the 
particular grid-based approach being adopted imposing a two-grid-cell 
minimum to the blastwave radius.  We speculated then that {\it any \rm} 
feedback scheme which distributed energy more efficiently on larger 
scales should result in flatter gradients.\footnote{Turbulence
driven by thermal instability can also be an efficient mechanism for
mixing metals, as elucidated upon in the excellent work of
\citet{Yang12}.}

At the time of this earlier work \citep{Pilk12a,Pilk12b}, we did not 
have a clear manner in which to quantify the above inference. We are now 
in a position to show, in a direct manner, the impact on the temporal 
evolution of abundance gradients for L$^\ast$ disc galaxies, when 
replacing the conventional feedback scheme employed for MUGS-g1536 and 
MUGS-g15785 with their now well-tested MaGICC analogs.  These enhanced 
feedback simulations (MaGICC-g1536 and MaGICC-g15784) were not available 
at the time of this previous work; combined with the aforementioned new 
empirical determinations of high-redshift in situ abundance 
gradients (most of which were also not available at the time of our 
initial study), this brief Research Note solidifies the more speculative 
conclusions we drew in \citep{Pilk12a}.

\section{Simulations}
We make use of two galaxies (\tt g1536\rm; \tt g15784\rm) drawn from 
the MUGS \citep{Stin10} suite of L$^\ast$ cosmological discs; these
two systems are both isolated and experienced relatively quiescent assembly 
histories, since redshift $z$$\sim$2. By avoiding strongly interacting
major mergers and/or close pairs, the comparison between simulation and
observation remains `like-with-like'; if we had not restricted ourselves
to such isolated systems, we would necessarily have had to consider
the impact that environment plays in
flattening gradients during periods of strong interaction 
\citep[e.g.][]{Rupk10,Few2012}. 
Realised with the SPH code \tt Gasoline \rm \citep{wadsley04}, two variants 
of each galaxy were analysed - one using `conventional' feedback (MUGS) and 
one using `enhanced' feedback (MaGICC).\footnote{It should be emphasised
that the assembly/merger histories for each MUGS-MaGICC `pair' (e.g., 
MUGS-g1536 and MaGICC-g1536) are identical; i.e., the differences discussed
in \S3 and \S4 are due to internal (e.g., feedback, star formation, etc.),
rather than external (e.g., merger history), processes.}
Full details, including the 
methodology associated with star formation and feedback, for MUGS, can 
be found in \citet{Stin10} and \citet{Pilk12a}. The MaGICC feedback 
prescription is outlined in a series of papers 
\citep{Broo11,Broo12a,Broo12b,Broo12c,Pilk12c,Stin12,Stin13}.

Briefly, MUGS-g1536 and MUGS-g15784 employ a thermal feedback scheme in 
which 4$\times$10$^{50}$ erg per supernova (SN) is made available to 
heat the surrounding ISM, while their MaGICC analogs use 
10$^{51}$~erg/SN.  The MUGS (MaGICC) simulations were realised with a 
\citet{Krou93} (Chabrier: \citealt{Chab01}) intial mass
function (IMF).\footnote{The MUGS runs 
assumed that the global metallicity Z$\equiv$O+Fe, while those of MaGICC 
assume Z$\equiv$O+Fe+C+N+Ne+Mg+Si; as such, the MUGS simulations 
underestimate the global metallicity by roughly a factor of two, and 
hence the impact of metallicity-dependent cooling \citep{Pilk12a}.} In 
the MaGICC runs, radiation energy feedback from massive stars is also 
included (in the $\sim$4~Myr prior to the appearance of the first 
Type~II SN from each star particle), albeit at an effective coupling 
efficiency $<$1\% \citep{Broo12b,Stin13}. For both MUGS and MaGICC, 
cooling is disabled for gas particles situated within a blast region of 
size $\sim$100~pc, for a time period of $\sim$10~Myr.  Star formation is 
restricted to regions which are both sufficiently cool ($<$10kK) and 
dense (MUGS: $>$1~cm$^{-3}$; MaGICC: $>$9~cm$^{-3}$). Metal diffusion 
\citep{shen10} is included in all runs.

To link the simulation nomenclature with their earlier appearances in 
the literature, MUGS-g1536 and MUGS-g15784 correspond to \tt g1536 \rm 
and \tt g15784\rm, respectively, in \citet{Stin10} and \citet{Pilk12a}, 
while MaGICC-g1536 corresponds to the `Fiducial' run in 
\citet{Stin13}.\footnote{MaGICC-g1536 is also essentially the same as 
SG5LR, as reported in our earlier work \citep{Broo12b}.}

\section{Abundance Gradients}
\begin{figure}[htb]
\begin{center}
\hspace{0.25cm}
\psfig{figure=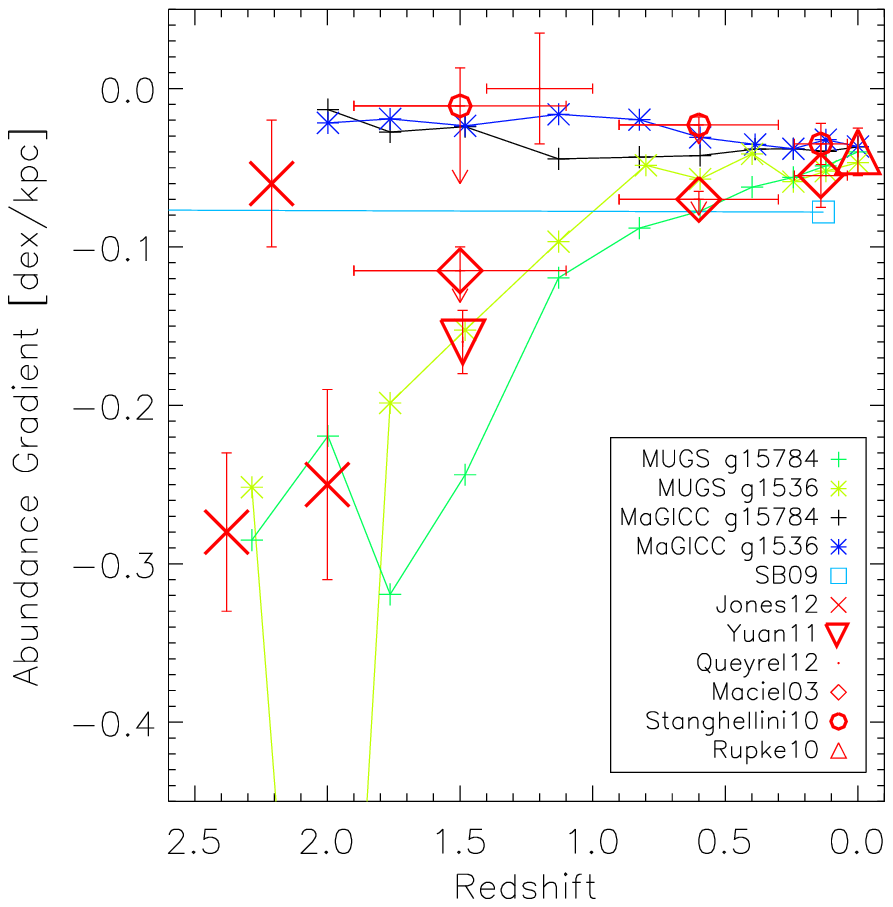,width=9.cm}
\caption{Temporal evolution of the predicted oxygen abundance gradients 
associated with four simulated L$^\ast$ disc galaxies - from the MUGS 
\citep{Stin10} and MaGICC \citep{Broo12b} suites - in addition to an 
analytical model (SB09: \citealt{Scho09}). Symbols correspond to 
empirical determinations of the abundance gradients in a sample of 
high-redshift lensed systems \citep{Yuan11,Jone12}, 
intermediate-redshift galaxies from MASSIV \citep{Quey12}, local discs 
\citep{Rupk10}, and the Milky Way \citep{Maci03,Stan10}.}
\label{fig1}
\end{center}
\end{figure}

In Fig~\ref{fig1}\footnote{Our Fig~\ref{fig1} here is a preferred update 
to Fig~5 of \citet{Pilk12a}.}, we show the evolution of the gas-phase 
oxygen abundance gradients for both sets of realisations (g15784: + 
symbols; g1536: $\ast$ symbols). Also included are the data sets of 
MASSIV \citep{Quey12}, \citet{Yuan11}, and \citet{Jone12}. These are, to 
date, the only high redshift results against which our simulations can 
be compared. We also include results from studies of local planetary 
nebulae by \citet{Maci03} (diamonds) and \citet{Stan10} 
(circles).\footnote{It is important to note that the `high-redshift' 
abundance gradients inferred from sub-types of planetary nebulae \it 
today \rm are strictly upper limits (as plotted), as any secular heating 
processes can really only flatten their in situ gradients to the 
values that we observe for them today; hence, the planetary nebulae data 
in Fig~\ref{fig1} are drawn with downward-facing arrows superimposed.}

The lower (`conventional' feedback) pair of simulated galaxies show 
significant flattening from $z$$\sim$2 to $z$$\sim$0, as previously 
described by \citet{Pilk12a}.\footnote{The $\sim$0.1$\rightarrow$0.3~dex
`deviations' in the flattening near redshift $z$$\sim$2 are transient
in nature, and due to periods of enhanced merger activity, as discussed
by \citet{Pilk12a} and \citet{Few2012}; this brief steepening of the
gradient, followed by a subsequent `return' to the global flattening
trend, is akin to the behaviour discussed eloquently by 
\citet{Rupk10}. Future work with a finer temporal output cadence
will be required to better quantify the timescale upon which the 
gradient `returns' to its global flattening trend.} 
Conversely, the upper pair show a 
dramatically different evolutionary sequence; the stronger feedback 
implemented within the MaGICC scheme, results in essentially flat 
gradients at high-redshift, with minimal steepening (rather than 
flattening) with time.  As the MaGICC feedback scheme re-distributes 
energy and re-cycled ISM material over much greater galactic scales
(via winds driving low-angular momentum inner disc and bulge gas
to the corona, where it then cools, falls back preferentailly
to the outer disc, 
and re-enters the star-forming region preferentially as an 
in-plane radial `flow'; the re-cycling pattern is described in more
detail by \citealt{Broo11,Broo12a})
such flat (and essentially temporally-invariant) gradients are consistent 
with our interpretation of the difference between the grid and 
particle-based simulations, as noted above, and in our earlier work 
\citep{Pilk12a}.  Specifically, the flatter gradients seen
in the MaGICC scheme at $z$$>$0 are due to the combined effect of 
(a) `metal re-cycling' via outflows (which re-distributes metals), 
and (b) `ISM re-structuring' via outflows (which re-distributes the 
ISM and hence radial star formation profile).

It is worth reminding the reader as to the predicted temporal evolution 
of the gradients from classical `analytical' chemical evolution models 
for the Milky Way \citep{Chia01,Moll05,Scho09}. As reported in 
\citet{Pilk12a}, the models of \citet{Chia01} and \citet{Moll05} show 
behaviour which is \it indistinguishable \rm from that seen in the 
stronger feedback MaGICC-g1536 and MaGICC-g15784 models.  The model of 
\citet{Scho09} is similar, in the sense of showing very little temporal 
evolution, albeit it remains somewhat steep at all times.

\begin{figure}[htb]
\begin{center}
\hspace{0.25cm}
\psfig{figure=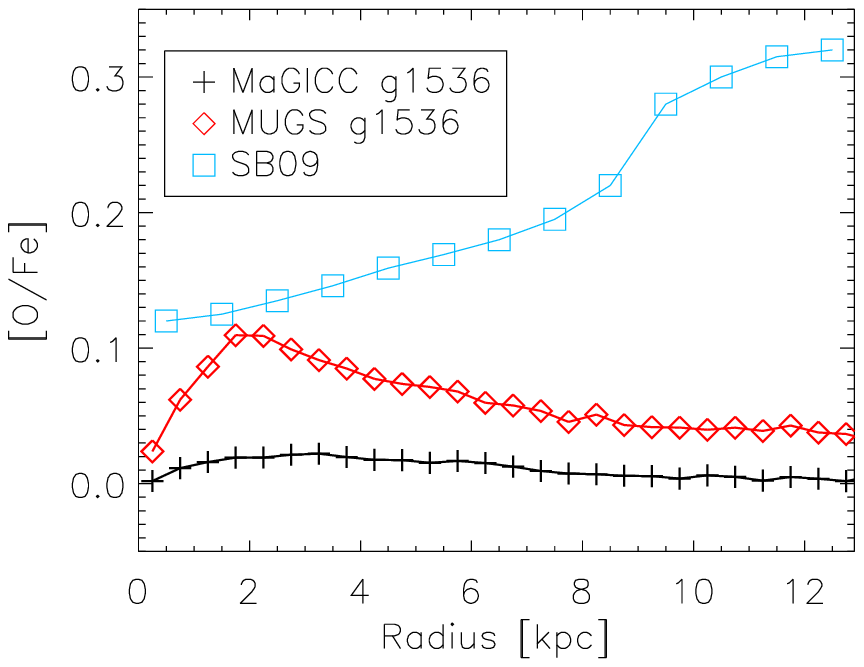,width=9.cm}
\caption{Predicted present-day, stellar mass-weighted, radial [O/Fe] 
gradients for the MaGICC \citep{Broo12b} and MUGS \citep{Stin10}
realisations of simulation \tt g1536\rm, compared with that predicted by 
an analytical model of the Milky Way (SB09: \citet{Scho09}).}
\label{fig2}
\end{center}
\end{figure}

It is worth asking how the transition from `conventional' MUGS feedback 
to stronger MaGICC feedback impacts on the inferred gradient in [O/Fe].  
Having addressed the former \citep{Pilk12b}, albeit briefly, we show in 
Fig~\ref{fig2} the mass-weighted stellar [O/Fe] gradients at redshift 
$z$=0 from MUGS-g1536, MaGICC-g1536, and the analytical model of 
\citet{Scho09}.  Both the MUGS and MaGICC realisations possess very flat 
gradients ($<$0.005~dex/kpc), similar to those observed by 
\citet{Sanc09}, although the uncertainties associated with inferring 
mass- or light-weighted [$\alpha$/Fe] gradients from integrated 
spectroscopy of face-on discs can be significant (see also 
\citet{Fenn06}).  In contrast with the MaGICC predictions, the 
analytical models of \citet{Scho09} predict {\it positive \rm} 
integrated light/mass gradients in [O/Fe], {\it within the star forming 
part of the disc\rm}, on the order of $\sim$$+$0.02~dex/kpc.

A characteristic of the model of \citet{Scho09}, relative to several 
classical models of galactic chemical evolution \citep{Chia01,Moll05}, 
is the inclusion of radial gas flows.\footnote{See also the radial flow 
model of \citet{Spitoni13}, for insightful commentary on the issue of 
radial flows in disc galaxies.} Such flows are also a 
natural outcome of our enhanced MaGICC feedback scheme \citep{Broo11}.  
While it can be challenging to infer the signal of $\sim$1$-$2~km/s 
flows, when superimposed upon a (say) $\sigma_r$$\sim$30~km/s velocity 
dispersion profile (both within the simulations and (especially) in 
nature \citep{Dame93}), we have attempted to do so.  Formally, for 
MUGS-g1536 (at redshift $z$=0), we find inward radial (cold) gas flows 
of $\sim$3~km/s ($\sim$1~km/s) within a $\pm$2~kpc thick annulus at 
13$<$$r$$<$17~kpc (7$<$$r$$<$8~kpc); for MaGICC-g1536, the radial flows 
(again, at $z$=0) are much larger: $\sim$12~km/s ($\sim$8~km/s) at the 
same galactocentric radii. A more detailed analysis of the temporal 
evolution of the gas flows will be required to disentangle the relative 
roles, within MaGICC, of re-cycling of the ISM over increasingly large 
galactic scales and the increasingly more substantial radial gas flows.

\section{Age Metallicity Relations}
\begin{figure*}[htb]
\begin{center}
\hspace{0.25cm}
\psfig{figure=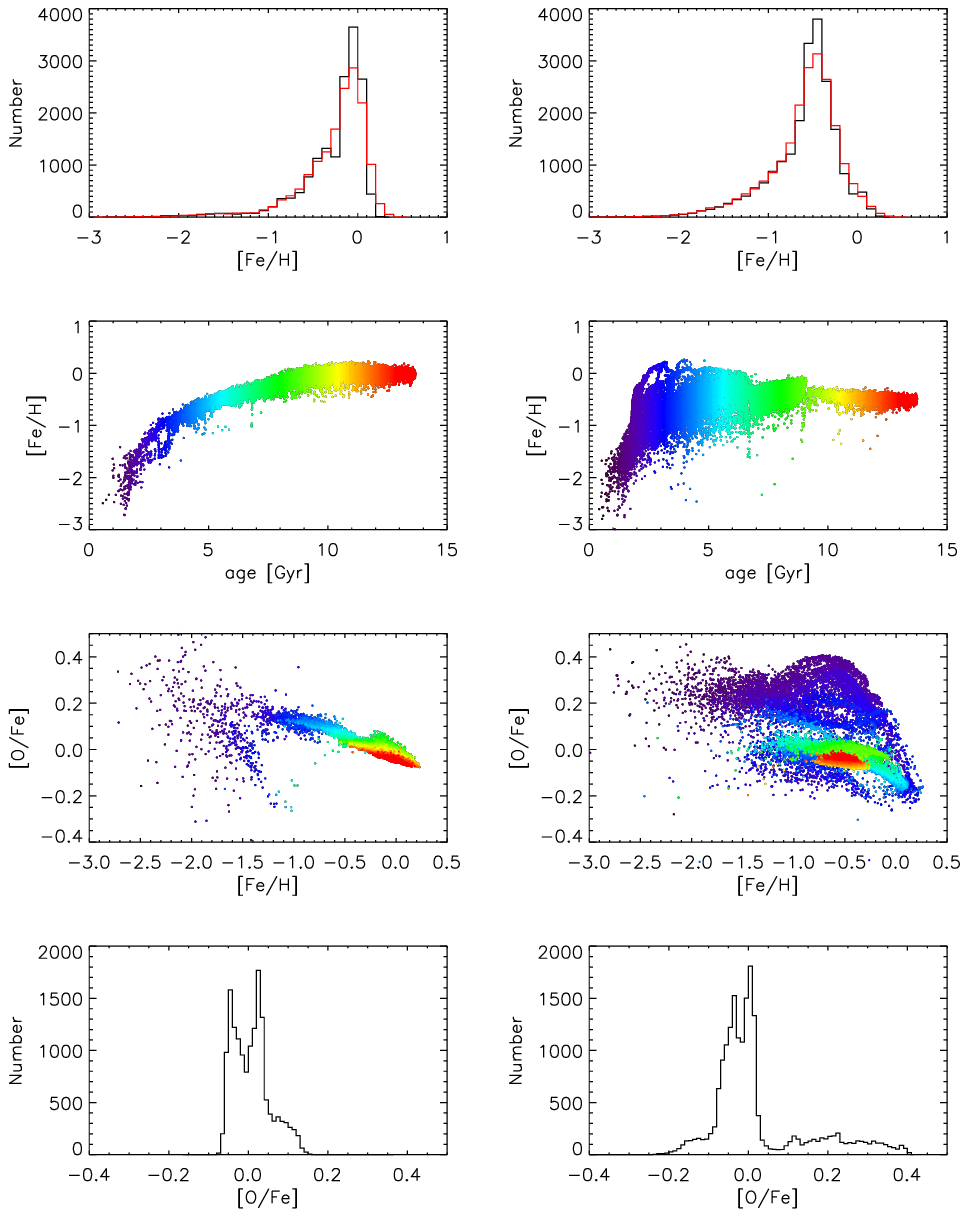,width=10.cm}
\caption{Predicted metallicity distribution functions (top row), 
age-metallicity relations (second row), [O/Fe]-[Fe/H] relations (third), 
and [O/Fe] distributions (bottom row), for `solar neighbourhood' regions 
of the MaGICC (\it left column\rm: \citealt{Stin13}) and MUGS (\it right 
column\rm: \citealt{Stin10}) realisations of simulation \tt g1536\rm.}
\label{fig3}
\end{center}
\end{figure*}

In Fig~\ref{fig3}, for MaGICC-g1536 (left column) and MUGS-g1536 (right 
column), we show their inferred local `solar neighbourhoods' stellar 
metallicity distribution function (MDF: top row), age-metallicity 
relation (AMR: second row), [O/Fe]-[Fe/H] distribution (third row), and 
[O/Fe] distribution function (bottom row). The solar neighbourhood is 
taken as the region 3$<$r$_{\rm d}$$<$3.5, where r$_{\rm d}$ is the 
radius in units of disc scalelength. While the impact of the enhanced 
feedback associated with MaGICC is readily apparent in the AMR and 
[O/Fe]-[Fe/H] planes, the effect is more subtle in the MDF and [O/Fe] 
distribution function.  The remarkably tight (effectively temporally 
invariant scatter) and correlated AMRs in the (analogous) simulated 
solar neighbourhood, for this L$^\ast$ realisation, is similar to that 
encountered in our earlier work on dwarf discs \citep{Pilk12c}.  At a 
given [Fe/H], within the solar neighbourhood, two (roughly) parallel 
loci in [O/Fe] co-exist (for [Fe/H]$>$$-$1) with a separation of
$\sim$0.1~dex (bottom left panel of Fig.~3); this separation is,
admittedly, smaller than that
seen in the solar neighbourhood of the Milky Way ($\sim$0.3~dex:
e.g., Fig.~3 of \citealt{Fuhrmann2008}), although in a qualitative
sense the behaviour is not dissimilar. 
An earlier detailed examination of the origin of these offset 
sequences (in a dwarf disc realisation) demonstrated significant 
parallels with said empirical sequences (drawing links with `thin' and 
`thick' discs, and highlighting the role of radial migration -
\citealt{Broo12c}); such behaviour, including offset loci in [O/Fe] for 
[Fe/H]$>$$-$1 is reflected in the L$^\ast$ realisations described here 
(see the bottom row of Fig~3, where the bimodal [O/Fe] distributions 
associated with [Fe/H]$>$$-$1 stars in the simulated solar 
neighbourhoods is apparent.

\section{Conclusions}

Negative abundance gradients at redshift zero provide a local boundary 
condition for galaxy evolution models, in support of inside-out disc 
growth. Importantly, we are seeing the beginnings of associated boundary 
conditions on the temporal evolution of said metallicity gradients, with 
recent in situ determinations of radial abundance gradients in 
typical star-forming and/or Grand Design spirals at redshifts 
2.4$<$$z$$<$1.0 \citep[e.g.][]{Yuan11,Quey12,Jone12}.  These 
developments are very much on the leading-edge of what can be done 
today; hence, the still contradictory nature of the results - i.e., 
whether or not abundance gradients steepen or flatten with time - should 
not be surprising.  Time (and additional high-redshift data) will 
certainly settle this observational issue; despite this, the power to 
use these in situ observations is enticing, and has led us to 
drive this effort to use their results to constrain the very uncertain 
nature of sub-grid physical energy feedback within models of galaxy 
evolution.

To this end, we have analysed a suite of simulated L$^\ast$ discs, 
realised with different feedback implementations. The enhanced feedback 
models of the MaGICC program \citep{Broo12b,Stin12}, which have been 
shown to be in agreement with a broad range of present-day empirical 
scaling relations, predict that gradients should only mildly steepen 
with time.  These relatively `flat' and temporally invariant abundance 
gradients result from feedback which distributes energy and re-cycled 
ISM material over large scales, coupled with stronger
radial gas flows.  These results are consistent with 
extant analytical models of galactic chemical evolution,
the inferred Milky Way gradient at
high-redshift by \citet{Stan10}, and in situ abundance
gradients at high-redshift, as determined by \citet{Quey12}. 
By contrast, the simulations 
which incorporated relatively weak feedback, without including early 
stellar feedback from massive stars prior to exploding as supernovae, 
results in metallicity gradients that are steep at high redshift 
(consistent with in situ abundance gradients at high-redshift 
determined by \citealt{Yuan11} and \citealt{Jone12}, 
and the inferred Milky Way gradient at 
high-redshift by \citealt{Maci03}). We do not wish to leave the
reader with any notions regarding the validity of MaGICC or 
MUGS feedback schemes based upon the in situ 
determination of abundance gradients at high-redshift; we are not
in a position to do so yet.  What \it is \rm true though is that
such empirical determinations possess a unique ability to 
constrain the uncertain nature of sub-grid feedback within
galaxy-scale hydrodynamical simulations. 

Enhanced feedback also results in significantly reduced scatter in the 
local stellar age-metallicity relation and, especially, the 
[O/Fe]-[Fe/H] relation. The local [O/Fe] distribution functions for our 
L$^\ast$ discs show clear bimodality, with peaks at [O/Fe]=$-$0.05 and 
$+$0.05 (for stars with [Fe/H]$>$$-$1); as noted in \S4, such a 
separation is qualitatively (if not quantitatively) similar to that 
seen in the solar neighbourhood of the Milky Way.  A detailed
analysis of both the age-metallicity relations and metallicity
distribution functions associated with these simulations will form
the basis of a future study. 

In light of the success of the MaGICC formulation for feedback 
at redshift zero 
\citep{Broo12a,Broo12b,Broo12c,Pilk12b,Pilk12c,Stin12,Stin13}, it 
may be tempting emphasise 
the consistency between classical models of chemical 
evolution, MaGICC-g1536, and MaGICC-g15784, with the empirical data of 
\citet{Quey12} and the planetary nebula work of \citet{Stan10}. Having 
said that, there are 
no a priori reasons to doubt the gradients 
inferred from lensed discs at high-redshift 
\citep{Yuan11,Yuan13,Jone12}, nor 
to dismiss the planetary nebulae work of \citet{Maci03}. Future 
observational campaigns, and detailed intercomparisons of the disparate 
planetary nebulae samples, will surely provide definitive and pivotal 
conclusions as to whether the somewhat flat and temporally invariant 
gradients predicted with the MaGICC feedback formulation stand the test 
of time, or whether the situation is more complicated (or at least 
varied) than we envision. Such observations are a unique and, until now, 
missing constraint/boundary condition on models of galaxy evolution.

\begin{acknowledgements}

BKG acknowledges the support of the UK’s Science \& Technology Facilities 
Council (ST/J001341/1).  KP acknowledges the 
support of STFC through its PhD Studentship programme (ST/F007701/1). 
The generous allocation of resources from STFC's DiRAC Facility
(COSMOS: Galactic Archaeology) is gratefully acknowledged. We also
thank the DEISA consortium, co-funded through EU FP6 project RI-031513 
and the FP7 project RI-222919, for support within the DEISA Extreme 
Computing Initiative, and the University of Central 
Lancashire’s High Performance Computing Facility.

\end{acknowledgements}

\bibliographystyle{aa}
\bibliography{Grad}

\end{document}